\documentclass[USenglish,oneside,twocolumn]{article}

\usepackage[utf8]{inputenc}
\usepackage[big]{dgruyter_NEW}
\usepackage[colorinlistoftodos]{todonotes}
\usepackage{comment}
\usepackage{multirow}
\usepackage{algpseudocode}
\usepackage{cleveref}
\usepackage{amsmath}
\usepackage{float}

\newcommand\norm[1]{\left\lVert#1\right\rVert}

\DOI{foobar}

\cclogo{\includegraphics{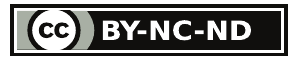}}
  
\begin{document}

  \author*[1]{Rafa G\'{a}lvez}

  \author[2]{Veelasha Moonsamy}

  \author[3]{Claudia Diaz}

  \affil[1]{imec-COSIC KU Leuven, E-mail: rafa@kuleuven.be}

  \affil[2]{Ruhr University Bochum, E-mail: email@veelasha.org}

  \affil[3]{imec-COSIC KU Leuven, E-mail: claudia.diaz@esat.kuleuven.be}

  \title{\huge Less is More: A privacy-respecting Android malware classifier using federated learning}

  \runningtitle{Less is More: A privacy-respecting Android malware classifier using federated learning}


  \begin{abstract}
    {In this paper we present LiM (`Less is More’), a malware classification framework that leverages Federated Learning to detect and classify malicious apps in a privacy-respecting manner. Information about newly installed apps is kept locally on users’ devices, so that the provider cannot infer which apps were installed by users. At the same time, input from all users is taken into account in the federated learning process and they all benefit from better classification performance. A key challenge of this setting is that users do not have access to the ground truth (i.e. they cannot correctly identify whether an app is malicious). To tackle this, LiM uses a safe semi-supervised ensemble that maximizes classification accuracy with respect to a baseline classifier trained by the service provider (i.e. the cloud). We implement LiM and show that the cloud server has F1 score of $95\%$, while clients have perfect recall with only $1$ false positive in $>100$ apps, using a dataset of 25K clean apps and 25K malicious apps, $200$ users and $50$ rounds of federation. Furthermore, we conduct a security analysis and demonstrate that LiM is robust against both poisoning attacks by adversaries who control half of the clients, and inference attacks performed by an honest-but-curious cloud server. Further experiments with MaMaDroid's dataset confirm resistance against poisoning attacks and a performance improvement due to the federation.}
\end{abstract}


  \journalname{Proceedings on Privacy Enhancing Technologies}
\DOI{Editor to enter DOI}
  \startpage{1}
  \received{..}
  \revised{..}
  \accepted{..}

  \journalyear{..}
  \journalvolume{..}
  \journalissue{..}

\maketitle



 
\section{Introduction}


Android dominates the mobile operating system market as the most popular choice amongst smartphone users. At the same time, this makes Android an attractive target for malware authors who want to infect as many devices as possible with malicious applications. Malware classifiers that leverage \emph{machine learning} (ML) techniques have been proposed to tackle this problem, showing varying degrees of success~\cite{arp_drebin_2014,MilosevicMachinelearningaided2017,SaracinoMADAMEffectiveEfficient2018,onwuzurike_mamadroid:_2019,mirzaei2019andrensemble,kim2018multimodal}. Often, to produce accurate classification results, machine learning models rely on access to a large and diverse set of features collected from user devices – which can be very revealing of the list of apps installed in each device~\cite{zhang2016,SaracinoMADAMEffectiveEfficient2018,arshad_samadroid_2018}. This raises privacy concerns: these models expose all users' private data to the centralized entity performing the classification, which may learn private information about the user; and motivates the need for decentralized, privacy-respecting malware classifiers for Android that can effectively protect users from malware infections without requiring them to expose private information to third parties.

ML solutions for malware classification can be grouped in three categories: (i) cloud-based, (ii) client-based and (iii) hybrid, i.e. a combination of (i) and (ii). In cloud-based solutions~\cite{onwuzurike_mamadroid:_2019, zhang2016}, the ML models are supplied with large sets of features that implicitly reveal users' installed apps, which constitute potentially sensitive data. In client-based solutions~\cite{arp_drebin_2014}, data is protected as the ML models are local and compute predictions on the device itself. This is however a resource-consuming approach that results in high false positives, as the models are not trained with sufficient new data as time goes by~\cite{SaracinoMADAMEffectiveEfficient2018}. Lastly, in hybrid solutions~\cite{SaracinoMADAMEffectiveEfficient2018,arshad_samadroid_2018} apps that are flagged as suspicious on the device are sent to the cloud for further analysis. Such solutions have a high number of false positives in the local models and reveal significant private data to the cloud. 

In general, it cannot be denied that ML algorithms are effective at detecting malware. However, of notable concern is the amount of information required for the ML models to produce good results and the associated repercussions on users' privacy. In particular, it can be observed that the larger the number of raw features centrally available for training and testing, the better the classification results. Additionally, as demonstrated by Song et al.~\cite{song2017}, ML models can memorize and leak detailed information about training datasets. This observation coupled with the fact that the list of apps installed on a mobile device may enable private inferences about the users' personal preferences, profile, behavior, etc. pose a serious threat to their privacy.



To address the aforementioned concern we investigate the following key question: \emph{How can we build a decentralized Android malware classifier that is privacy-preserving?}

In this paper, we present LiM -- a hybrid solution that leverages the power of Federated Learning (FL) to provide a malware detection solution that has high classification performance while respecting user privacy. LiM can classify all the apps installed on users' devices regardless of whether they are obtained from an app store or other sources, allowing users to detect malware without relying on the hosts of app stores (e.g. Google for Play Store) for malware classification services. LiM reduces the dependency of users on app stores in a way that benefits both privacy and malware detection performance.

State of the art FL models~\cite{yao2019, yang2019federated, milad_nasr_comprehensive_2019} allow users to keep their testing data locally while the learning process is done collaboratively to improve performance, i.e. users train their client models with a supervised algorithm, while a service provider aggregates the parameters of all models. LiM extends the traditional FL technique to the safe semi-supervised ML paradigm~\cite{kairouz_advances_2019}, enabling the application of FL in settings where users do not have local access to ground truth, as is the case with malware classification. Safe semi-supervised models combine labeled data available to the cloud server with unlabeled data available to clients. The cloud server trains fully-supervised models and shares them with clients, who in turn retrain with their unlabeled local data without introducing a performance penalty.

We validate the design by implementing a prototype of LiM and evaluating its performance and its security against poisoning and inference attacks. We carry out experiments using a dataset of 25K malware apps and 25K clean apps, simulating federations of $200$ clients over $50$ rounds. The results show that the cloud can reach $95\%$ F1 score, and clients has as few as $1$ false positive. Additionally, faced with a strategic adversary that controls $50\%$ of the clients and whose goal is to perform a poisoning attack, the remaining honest clients are able to correctly identify the targeted, poisoned app. Moreover, we demonstrate that the cloud server is unable to infer whether a specific app was installed by any of the clients, making LiM resistant against membership inference attacks. Finally, we validate LiM's capability to learn from the federation using MaMaDroid's dataset~\cite{onwuzurike_mamadroid:_2019}, as well as its resistance against poisoning attacks.




\noindent\textbf{Our contributions}:
\begin{enumerate}
  \item We present a first, comprehensive design and implementation of a privacy-respecting Android malware classifier.
  \item We demonstrate an effective way to combine FL and safe semi-supervised ensemble learning to enhance malware detection accuracy and privacy at the same time.
  \item We conduct a security analysis to illustrate the robustness of LiM against poisoning and inference attacks.
  \item In the spirit of open science, we make our code available at \url{https://git.sr.ht/~rafagalvez/lim-python}.
\end{enumerate}

\noindent\textbf{Terminology}. To improve readability of the remaining sections, we provide our working definitions of key terms used throughout the paper. 
\begin{itemize}
\item \textbf{Client} refers to the ML model that resides locally on the user's mobile device. 
\item \textbf{Cloud} refers to the global ML model which is present at the service provider's side.
\item \textbf{SAFEW} is an ensemble classifier composed of a set of individual base learners and their weights.
\item \textbf{Baseline classifier} is a standalone classifier whose performance has to be met by all base learners (combined) as a lower bound; also referred to as \emph{baseline}.
\end{itemize}

\noindent\textbf{Roadmap}. The rest of the paper is organized as follows: in section~\ref{sec:background}, we provide background knowledge about Federated Learning, semi-supervised learning and Android malware classification. Section~\ref{sec:threat-model} describes the threat model of LiM. In section~\ref{sec:semi-supervised-fl}, we elaborate on safe semi-supervised FL and how it is implemented in LiM. Section~\ref{sec:architecture} provides the details of the LiM architecture and its associated building blocks, followed by the empirical results and security analysis in section~\ref{sec:evaluation}. Section~\ref{sec:discussion} presents a discussion based on the empirical results and avenues for future work together with relevant related work in section~\ref{sec:relatedwork} and concluding remarks in section~\ref{sec:conclusion}.

\section{Background}\label{sec:background}

Federated Learning (FL), also referred to as `Collaborative Learning', is a technique where an ML algorithm is iteratively trained in a distributed setting with a client-server architecture~\cite{konevcny2015federated}: a large number of clients contribute locally-computed learning model parameters to a service provider (e.g. a cloud server) that aggregates those client parameters to compute updated parameters for the federated model, which is in turn sent back to clients for the next iteration. Each iteration, or \emph{round}, improves the performance of the client models thanks to 1) the newly available local data that can further refine training, and 2) the aggregated parameters shared by the cloud server, which improve the model based on learning done by the ensemble of all clients. 

FL offers privacy advantages compared to purely centralized models where the service provider collects raw input data from all clients in order to train the model. FL models keep data local to the clients and instead share model parameters.
However, the client-server architecture creates opportunities for adversaries that compromise a fraction of clients or the service provider to impact performance and privacy guarantees. On the one hand, poisoning attacks~\cite{BagdasaryanHowBackdoorFederated2018} may harm the performance of the federated model if adversarial clients submit maliciously crafted parameters to the server. On the other hand, client-provided parameters may be exploited by a curious server that conducts inference attacks~\cite{melis_exploiting_2019} to identify the training examples used by clients. In this paper, we tackle both threats by making use of safe semi-supervised learning to: 1) enable the cloud server to exclude poisoned parameters by using locally available data, and 2) allow clients to share hyper parameters rather than the parameters of individual classifiers, which cannot be exploited to infer training inputs.

LiM uses FL for malware classification. One limitation of current FL solutions is that they rely on supervised learning, requiring clients to have access to ground truth for the local training process~\cite{kairouz_advances_2019}. While applications such as predictive typing can benefit from this approach (since the client can locally test predictions against actual inputs from the user), this assumption does not hold in malware classification use cases, where clients do not have local knowledge of the ground truth.

LiM solves this problem using a safe semi-supervised learning algorithm that allows clients to use unlabeled data for training, while guaranteeing that the performance will be at least as good as that of a baseline classifier.

\subsection{Safe semi-supervised learning}
\label{sec:safe-ssl}

Semi-supervised learning (SSL) uses both labeled and unlabeled data to train a classifier. Labeling data can be expensive, while collecting and learning labels from raw data has become easier with the commoditization of internet access and the plethora of apps installed on smartphones. One of the main challenges for the success of an SSL algorithm is to ensure it indeed learns useful information from unlabeled samples, as there is no ground truth for the algorithm to verify its predictions for those samples.

\textbf{Safe SSL} addresses this concern by ensuring that a minimum baseline performance is always achieved, i.e. that unlabeled information does not worsen the performance of another (fully supervised) classifier. A well-performing strategy to achieve safe SSL is to use an ensemble of learners that, combined through a set of learned weights, are likely to outperform the baseline model \cite{li_towards_2014,li_towards_2019}.

An example of this kind of classifier is SAFEW~\cite{li_towards_2019}. Its goal is to combine a set of classifiers, called \emph{base learners}, through a set of weights $\alpha_{i}, i \in [1, b]$ that leads the ensemble to perform always better than a given baseline classifier. SAFEW achieves this goal assuming the ground truth label assignment of the unlabeled data $f^{*}$ can be realized as a convex combination of the predictions from the base learners \emph{b}, i.e. $f^{*} = \sum_{i=1}^{b}\alpha_{i}\textbf{f}_{i}$~\cite{li_towards_2019}. Using this assumption, SAFEW can compute the error of the baseline classifier $\mathbf{f}_{0}$ and final $\mathbf{f} \in \{+1, -1\}$ predictions with respect to the ground truth $f^{*}$ using the loss function $l$, and make $l(\mathbf{f},f^{*})$
as small as possible compared to $l(\mathbf{f}_{0}, f^{*})$ even in the worst case, maximizing the minimum difference as shown in equation~\ref{eq:safew-maximin}.

\begin{equation}
  \label{eq:safew-maximin}
  \max_{\mathbf{f} \in {\{+1, -1\}}}{
    \min_{\mathbf{\alpha}}{
      l(\mathbf{f}_{0}, \sum_{i=1}^{b}\alpha_{i}\mathbf{f}_{i})
      -
      l(\mathbf{f}, \sum_{i=1}^{b}\alpha_{i}\mathbf{f}_{i})
    }
  }
\end{equation}

SAFEW does not impose restrictions on which classifiers to use as base learners to predict $\mathbf{f}_{i}$: results do not depend on the amount of base learners but on their quality, and different learning algorithms can be used, for instance Support Vector Machines together with Random Forests. Similarly, there is no a priori restriction on the loss function that can be used as $l$. However, to reduce computation time and find optimal weights $\mathbf{\alpha}^{*}$, Li et al.~\cite{li_towards_2019} show that using the hinge loss function turns equation~\ref{eq:safew-maximin} into a convex optimization problem that can be solved with common optimization packages like the ones supported by CVXPY~\cite{diamond_cvxpy_2016,agrawal_rewriting_2018}. Prior knowledge can be embedded as constraints in this problem formulation in order to enhance the information extracted from the unlabeled data. The final predictions $\bar{f}$ can then be obtained through equation~\ref{eq:safew-predictions}.
\begin{equation}
  \label{eq:safew-predictions}
  \bar{f} = \sum_{i=1}^{b}\alpha_{i}^{*}\mathbf{f}_{i}
\end{equation}

The solution proposed by SAFEW assumes a centralized setting where labeled and unlabeled data rest in the same place. Such scenario requires users to share data with the service provider, potentially leaking sensitive information. LiM addresses this gap by federating SAFEW, letting users keep their data local while sharing only the weights that enable SAFEW to improve performance over baseline. Moreover, LiM also aims to be resistant against membership inference and poisoning attacks, addressing the increasingly important security and privacy concerns that arise from the use of advanced machine learning in the wild.

\subsection{Android}
\label{sec:org2c06dab}

\subsubsection{Android manifest file}
In the Android OS, apps are distributed as Android application package (APK) files. These files are simple archives which contain bytecode, resources and metadata. A user can install or uninstall an app (the APK file) by directly interacting with the smartphone. When an Android app is running, its code is executed in a sandbox. In theory, an app runs isolated from the rest of the system, and it cannot directly access other apps' data. The only way an app can gain access is via the mediation of inter-process communication techniques made available by Android. These measures are in place to prevent the access of malicious apps to other apps' data, which could potentially be privacy-sensitive.

Since Android apps run in a sandbox, they are subject to restrictions on the usage of shared memory and most system resources. The Android OS provides an extensive set of Accessible Programming Interfaces (APIs) that allows access to system resources and services. In particular, the APIs that give access to potentially privacy-violating services (e.g., camera, microphone) or sensitive data (e.g., contacts) are protected by the Android Permission System~\cite{androidPermission}. Developers have to explicitly mention the permissions that require user's approval in the \texttt{AndroidManifest.xml} file (hereon referred to as the Manifest file). 

Besides permissions, the Manifest file also includes information about the app components~\cite{appfundamentals}, such as activities, services, broadcast receivers and content providers. An \emph{activity} is the representation of a single screen that handles interactions between user and apps. \emph{Services} are components that run in the background of the OS to perform long-running operations while a different application is running in the foreground. \emph{Broadcast receivers} respond to broadcast messages from other applications or the system. They allow an app to respond to broadcast announcements outside of a regular user flow. A \emph{content provider} manages a shared set of app data and stores them in the file system. It also supplies data from one app to another on request.

It is worth noting that the information present in a Manifest file is not obfuscated and can be extracted via static analysis. It is in the app developer's best interest to not obfuscate the file as it would result in breaking the functionalities of the app, rendering it useless. In section~\ref{sec:evaluation}, we provide further details about the features used by our proposed classifier, LiM, to conduct malware detection. 

\subsubsection{Android malware classification}

There are several proposals for ML classifiers that can detect malicious APKs targeting Android. We divide them in three categories: centralized, local and hybrid.

Centralized approaches use a cloud classifier to predict if an app is malicious or clean. Cloud-based approaches can accurately predict big testing datasets thanks to the advanced feature engineering a cloud infrastructure can handle \cite{onwuzurike_mamadroid:_2019}. Both static and dynamic analysis can be performed, for e.g. such as taking into account the API call graphs of the apps and behavioral characteristics of an app during execution.

Local approaches install an already-trained classifier on the user's device. Due to the constrained resources available to the classifier, the feature set and the detection algorithm must be considerably more lightweight than in centralized approaches \cite{arp_drebin_2014}. Lightweight dynamic analysis can be performed together with static analysis (for e.g. features from the Manifest file).

Hybrid approaches combine local and cloud models. A first screening of the app is performed on the client device itself using a lightweight feature set, and if necessary more features are collected and sent to the cloud to verify the prediction \cite{SaracinoMADAMEffectiveEfficient2018,arshad_samadroid_2018}.

Existing centralized and hybrid solutions achieve good performance but expose user inputs to the service provider, enabling potentially sensitive inferences. Local approaches do not expose client data, but suffer from poor performance. In constrast, LiM's performace is comparable to that of centralized and hybrid solutions, while its privacy protection is as in local approaches, obtaining the best of both worlds.

\section{Threat model}
\label{sec:threat-model}

LiM's functionality is to classify malware in a privacy-preserving manner, using base learners trained with data that never leaves the client device and yet contributes to improving the accuracy of classification for all clients. In our threat model, we distinguish two types of adversaries depending on their attack goals:
\begin{enumerate}
\item The goal of \emph{Adversary 1} is to compromise \textbf{integrity}: poison the federated learning with maliciously crafted inputs in order to trigger specific apps to be misclassified, as presented in~\cite{BagdasaryanHowBackdoorFederated2018}.
\item The goal of \emph{Adversary 2} is to compromise \textbf{privacy}: infer private information about the training data of clients (list of installed apps), as presented in~\cite{melis_exploiting_2019}.
\end{enumerate}

Adversary 1 targets the core functionality of the malware classifier, i.e. bypass the malware detection mechanism. To achieve that, the adversary modifies the classification model so that it misclassifies a specific malicious application. In a federated setting, the adversary may control a subset of users (at most 50\%) who submit maliciously crafted model parameters to the federation. Further, we assume that the adversary controls the malicious application, which can be tweaked so that its features resemble clean apps and confuse the model into misclassification. In terms of trust model, we assume that non-malicious users compute and submit correctly computed parameters, and that the cloud server correctly follows the learning process.

Adversary 2 aims to compromise user privacy by inferring information about the apps that users have installed on their devices, e.g. the app names, categories, usage patterns, etc. In a federated setting, we are interested in a passive global adversary, as described in~\cite{milad_nasr_comprehensive_2019}, that has access to the cloud server's data including all client model parameters uploaded to the cloud. The adversary examines these individual client models to try to infer information about which apps users have installed. Note that LiM relies on information that is statically extracted from app manifests and is thus only concerned with app installs, while inferences on app usage after installation is out of scope. 

Both adversaries have white-box access to the cloud model (including architecture, feature set and hyper-parameters) in each federation round. Adversary 2 also has white-box access to the models of all clients in each round, while Adversary 1 does not know the hyper-parameters of the models of honest users.

\section{Safe semi-supervised federated learning}
\label{sec:semi-supervised-fl}
Traditionally, FL employs a decentralized approach to train a neural model. Instead of uploading data for centralized training, clients process their data locally and share the resulting model updates with the service provider. Such distributed approach has been shown to work with unbalanced datasets and data that is not independent or identically distributed across clients. 

FL's success is, however, dependent on properly labeled data that can be used to train supervised learning models. Considering the scenario of malware classification, we cannot rely on users assigning correct labels to their data, as it cannot be guaranteed that they can correctly identify malicious apps. 

LiM proposes a novel approach to address this challenge: a semi-supervised method that allows FL to train local models without supervision. We assume that labeled data is only available to the cloud, while clients use their unlabeled samples to update the parameters of their local semi-supervised model.

Furthermore, we leverage the practical benefits of safe semi-supervised learning (SSL), as described in section~\ref{sec:safe-ssl}, to ensure that models trained by the clients are useful, i.e. that they provide a minimum baseline performance and do not introduce noise (via incorrect labels) in the federated model. 

\begin{figure*} [htb]
  \centering
  \includegraphics[width=0.9\linewidth]{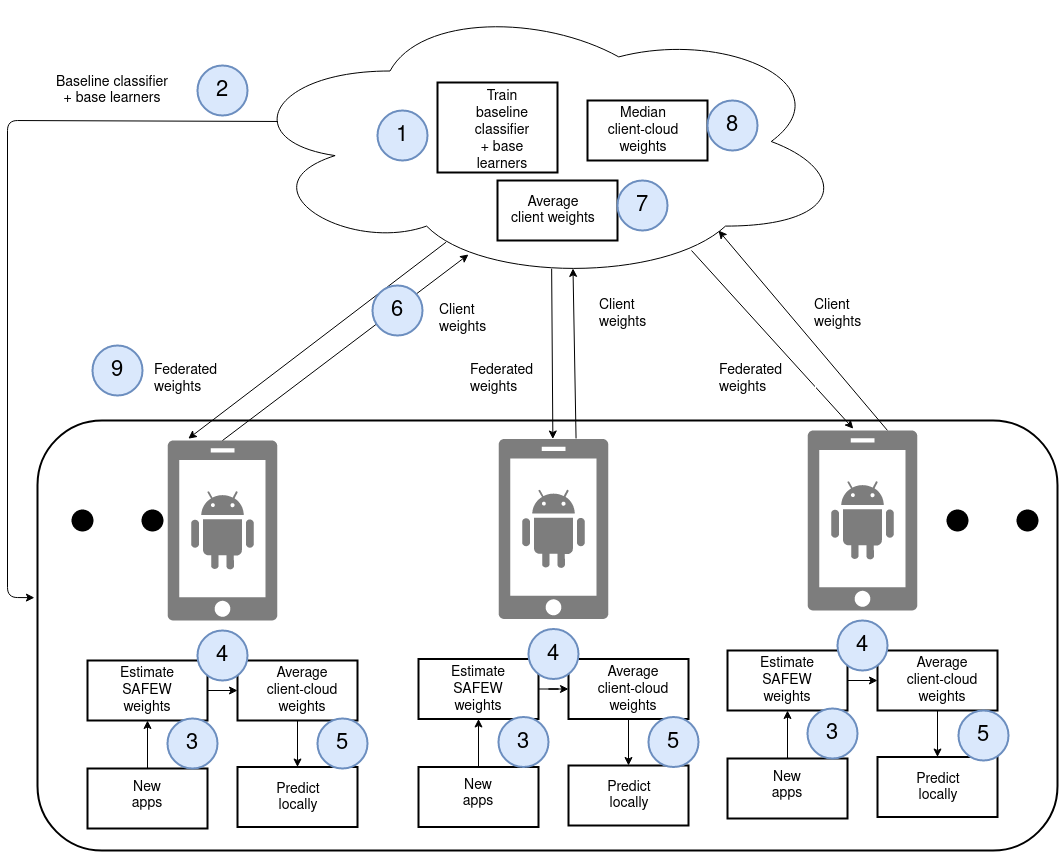}
  \caption{LiM architecture -- the cloud and the clients all train their own SAFEW model, and weights are aggregated twice to distribute the information of the testing datasets of the users.}
\label{fig:lim-architecture}
\end{figure*}

In LiM, the federation happens across the weights of the base learners, which clients estimate using their unlabeled testing datasets. The cloud server collects all client weights and aggregates them in a similar fashion as it would do with the weights of e.g. a deep neural network (DNN). It is important to note that the number of base learners is much lower than the number of neurons in a DNN -- LiM further compresses client data by taking advantage of the training process that the cloud performs on the base learners. We see this feature as a defense mechanism against \emph{privacy attacks} (cf. section~\ref{sec:threat-model}), as client updates will not be sparse anymore.

Note that in the case of LiM, the service provider plays a greater role than in the classical, supervised FL to compensate for the lack of ground truth in the clients. The provider also selects the architecture and the feature sets of the different base learners, as well the one to be used as baseline classifier. LiM provides protection against \emph{integrity attacks} (cf. section~\ref{sec:threat-model}) by comparing the weights contributed by clients with those generated by the provider through its own unlabeled dataset.

\section{LiM architecture}
\label{sec:architecture}

We assume that the service provider (cloud server) has access to a ground truth (labeled) dataset of malware and clean apps, and a testing (unlabeled) dataset. At the client-side, users locally scan their installed apps to identify malware. LiM can be incorporated in the package installer of an Android OS and executed as a privileged background service. LiM uses the SAFEW classifier described in section~\ref{sec:safe-ssl} to implement the scheme explained in section~\ref{sec:semi-supervised-fl}. 
\vspace{2mm}\\

\noindent{\textbf{Round 0 of FL}}: The process is depicted in figure~\ref{fig:lim-architecture}. First, the server uses (step 1) the labeled dataset to train a baseline classifier and a set of base learners, and the unlabeled dataset to estimate weights for the base learners. Clients receive (step 2) the baseline classifier and trained learners in order to estimate (step 3) their local SAFEW weights using their own testing dataset (i.e. their installed apps). Clients then compute average weights (step 4) and use them to classify their installed apps (step 5). Users then complete the federation round by sending their locally computed weights to the cloud (step 6). The cloud averages all the client weights (step 7) and then further averages that value (step 8) with the weights of its own SAFEW, computed in step 1. Finally, the cloud sends the updated federated weights to the clients to initiate a new round of federated learning (step 9). 
\vspace{2mm}\\

\noindent{\textbf{Round 1 of FL (and beyond)}}:
Once Round 0 is completed, steps 1-2 are optional: the service provider can re-apply them at any point in the subsequent federation rounds. In our evaluation, we did not conduct experiments that consider such updates. Steps 3-9 are applied as described in Round 0, even if some users do not install new apps. We account for this possibility in our evaluation.

Note that in LiM, the service provider does not send its own weights in the initial round (i.e. Round 0) to prevent an adversary from using this information to craft an integrity attack (cf.section~\ref{sec:threat-model}), and to give clients' inputs greater influence in the federated weights. In later rounds, new federated weights are computed by using both client and cloud weights. The provider computes the mean of all clients' weights and averages that with its own locally computed weights. The design behind this construction is purposefully conservative to counter integrity attacks where malicious inputs degrade overall classification accuracy -- guaranteeing the key functionality of a decentralized malware classifier.

\vspace{2mm}

LiM uses SAFEW as the local classifier that clients train in order to predict if their installed apps are malicious. To take advantage of the optimizations explained in section~\ref{sec:safe-ssl}, the individual SAFEWs use the hinge loss function to estimate their weights from unlabeled data. Regarding the choice of SAFEW base learners, it is possible to choose any base classifier as the optimization algorithm of SAFEW only uses the predictions of base learners to compute their weights. There are two main criteria to keep in mind when selecting alternative learners:
\begin{itemize}
\item They must provide a reasonable performance on their own, e.g. over 90\% F1 score.
\item Their predictions must complement each other, i.e. the learners must be heterogeneous. If there is one clearly strong learner, SAFEW will just always copy its predictions (i.e. its weight will be 1).
\end{itemize}

Additionally, domain knowledge can guide the aforementioned selection process, and it is possible to further constrain the set of possible weights $\alpha_{i}$ (cf. section~\ref{sec:safe-ssl}) to reflect, e.g. the confidence that the designer has on each learner relative to others. It is, however, important to note that LiM does not need a lot of expert knowledge for its setup. Weights can be learnt from data without any prior domain knowledge, and there are no assumptions on the distribution of the testing dataset, i.e. no prior knowledge on the proportion of samples per class.


In section~\ref{sec:evaluation}, we compare the performance of standard learners to find which combination is the most effective for malware detection.

\section{Evaluation of the system}\label{sec:evaluation}

We empirically evaluate LiM as a federated malware classifier in a setup with a server and 200 clients iterating over 50 federation rounds. In our evaluation, the clients run as a parallelized (across clients) Python program on a 4 Intel(R) Core(TM) i5-4590 CPU @ 3.30GHz cores using $7.5$ GiB of RAM. The goal of the evaluation is to show how LiM performs using different configurations for the individual SAFEWs, as well as its evolution across rounds with respect to the different baseline learners and local (i.e. non-federated) SAFEWs.

Specifically, we set up LiM to use the following learners: $k$ Nearest Neighbours with number of neighbors $n = 3 (kNN n3)$, Logistic Regression with regularization parameter $c = 1 (LR c1)$, Random Forests with number of decision trees $n=50 (RF n50)$, $100 (RF n100)$ and $200 (RF n200)$, and Linear SVM with regularization parameter $c = 1 (SVM c1)$. All base learners are used in each experiment, rotating the baseline classifier role across them. Clients only use configurations that the cloud has vetted as safe, i.e. where the cloud SAFEW outperformed or matched the performance of the baseline classifier.

We perform 4 experiments per configuration using the Top $100$, $200$, and $500$ features (i.e. 12 experiments per configuration). Half of the experiments are done with $50\%$ of adversarial clients to evaluate resilience against integrity attacks.

The rest of this section describes the datasets used and the results of the performance and security evaluation of LiM. 

\subsection{Datasets}\label{subsec:datasets}
We use the AndroZoo dataset~\cite{AllixAndroZooCollectingMillions2016}  to obtain 25K clean
apps, which were selected from the top 3 most popular stores (Anzhi, Appchina and Google Play Store) as of October 2018. As for the malware samples, we collected 25K samples from the Android Malware Genome project~\cite{zhou2012dissecting} and the Android Malware Dataset project~\cite{wei2017deep, li2017android}. We pick the latest version of apps, removing duplicates within the same store. It is possible that the same app is published in two different stores as different versions, but we consider they are effectively different apps as developers may include different functionality for particular stores.

For each app, we extract the features from the Manifest file, thus following the recommendation proposed by the authors of Drebin in~\cite{arp_drebin_2014} -- a key related work in Android malware detection and feature extraction. While dynamic analysis can provide greater performance, it is both more resource intensive and easily obfuscated than static analysis. We then transform the statically extracted features into a vector of numerical binary values indicating the presence of a feature (e.g. a specific permission) in the Manifest file of an app. Finally, out of 370K features we select the top 100, 200, and 500 features ranked according to their chi-squared scores with respect to the ground-truth class. Table~\ref{tab:feature-categories} depicts the number of features and their corresponding feature types. To maintain consistency, we use the same categories as presented in~\cite{arp_drebin_2014}.

\begin{table}
\centering
\caption{Number of features per feature types. In the Top 500, there are 55 features that belong to two categories: 54 of them can be declared as permissions or hardware components, the 55th (com\_facebook\_facebookcontentprovider) can be either an activity or a content provider.~\label{tab:feature-categories}}
\begin{tabular}{lccc}
\toprule
  Types of features & \multicolumn{3}{c}{Number of features}\\
\cline{2-4}
{} &  Top 100 &  Top 200 &  Top 500 \\
\midrule
declared permissions &   65 &   73 &   97 \\
activities           &   19 &   80 &  232 \\
services             &    3 &   13 &   71 \\
intent filters       &    0 &    0 &    0 \\
content providers    &    1 &    3 &   10 \\
broadcast receivers  &    6 &   25 &   85 \\
hardware components  &   24 &   24 &   24 \\
\bottomrule
\end{tabular}
\end{table}

Out of the 50K apps, we randomly sample a training set of 10K apps for training the baseline and base learners. The cloud SAFEW testing set has 32K apps, and the overall client testing set has 8K apps. In order to facilitate the learning process between cloud and clients, both testing data sets have an overlap of \textasciitilde1K apps.

We simulate several rounds of federation as described in section~\ref{sec:architecture}. In the first round, clients have a set of 96 preinstalled apps, which have been extracted from an Android Pie emulator and whose manifests have at least one permission. In later rounds, the clients install up to 5 apps drawn randomly from the client testing dataset, using a binomial distribution with bias 0.6 to randomize the number of apps. Each app will be a malware sample with probability 0.1. Moreover, to model the fact that users install popular apps much more frequently than others, we create two sets of 50 apps based on the presence of popular features in the malware and clean datasets, and make clients draw apps from these sets with probability 0.8.

\subsection{Performance evaluation}\label{subsec:performance-evaluation}

\subsubsection{Performance metrics}
\label{sec:performance-metrics}

To evaluate LiM in a realistic setting where users encounter many more clean apps than malicious ones (i.e. where classes are highly imbalanced) we use the F1 score, computed as F1 $= 2 * (precision * recall) / (precision + recall)$. Precision measures how many positive predictions (true positives + false positives) were actual positives (true positives), while recall measures how many actual positives (true positives + false negatives) were classified as positive (true positives). Since users typically install many more clean apps than malicious apps, it is easy for LiM to achieve high recall by predicting many positives at the expense of precision, which is not accounted for in other popular metrics like accuracy.

In order to better understand this balance between precision and recall across SAFEW configurations, we also report the raw number of false positives. We argue that, irrespective of whether the F1 score of two versions of LiM using different configurations of base learners is very similar, end users are rather sensitive to small differences in the number of false positives. More concretely, if a malware classifier repeatedly flags clean apps as malicious, then this will significantly affect user experience. 

\subsubsection{Performance results}
\label{sec:performance-results}

In the cloud, we observe that LiM matches the F1 score of the centralized SAFEW most of the time, and that the number of false positives is the second lowest using RFn50 with 500 features (cf. table~\ref{tab:performance-cloud} from the appendix). Regardless of the configuration used, the F1 score remains around between 94\% and 96\%.

\begin{figure}[hbt!]
  \centering
  \includegraphics[width=0.5\textwidth]{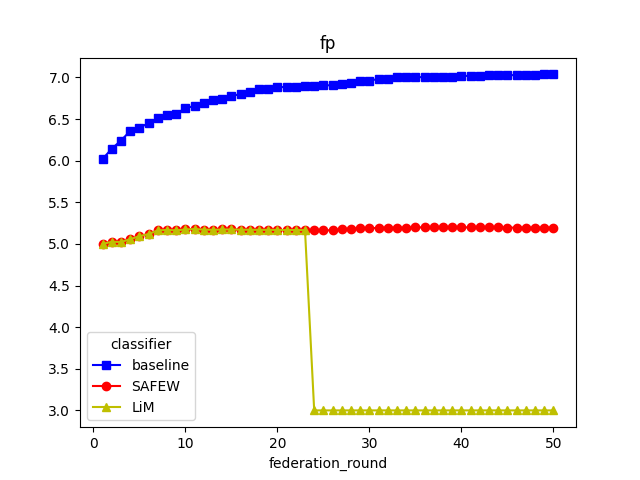}
  \caption{False positives of 200 clients with 500 features, using kNN (n=3) as baseline classifier. LiM reduces the number of false positives to 3, while both SAFEW (i.e. no FL) and baseline have 5 and 7, respectively. \label{fig:client_fp_intervals}}
\end{figure}

In figure~\ref{fig:client_fp_intervals}, we can see how this performance gain comes from the reduction of false positives. The evolution shows that learning from the federation allows LiM to drop the number of false positives from 5 (same as SAFEW) to 3, whereas baseline KNN misclassify more clean apps as we advance through the FL rounds.

Table~\ref{tab:performance-clients} shows the F1 score and the number of false positives (FP) of clients using different baseline classifiers and sets of base learners, and averaged across experiments. The best performance, i.e. F1 score of 73.7\% is achieved using RFn50 as baseline classifier and 200 features, thanks to the low number of false positives. Using kNN as baseline classifier gives us a high F1 score of 73.2\% using 200 features.

\begin{table*}[h!]
\centering
  \caption{Comparison of client average performance across LiM, SAFEW, and different baselines. Experiments are carried out for 50 rounds of federation using the Top 100, 200 and 500 features which were selected based on the chi2 test\label{tab:performance-clients}.}
\begin{tabular}{l|c|ccc|ccc}
\toprule
Baseline & \#Features & \multicolumn{3}{c|}{FP (raw values)} & \multicolumn{3}{c}{F1 (\%)} \\ \hline
       &  & Baseline &  SAFEW  &  LiM & Baseline &    SAFEW  &  LiM \\ 
\midrule
  KNN n3 & 100 &    8.9 &  5.6 &  2.4 &    36.4 &  46.5 &  66.5 \\
       & 200 &    7.6 &  4.1 &  1.6 &    39.8 &  53.8 &  \textbf{73.2} \\
       & 500 &    6.7 &  3.6 &  3.0 &    37.9 &  55.2 &  58.5 \\
LR c1 & 100 &    5.1 &  5.4 &  4.0 &    47.5 &  48.5 &  54.3 \\
       & 200 &    3.6 &  3.6 &  3.6 &    55.2 &  55.4 &  55.0 \\
       & 500 &    4.2 &  4.2 &  4.2 &    52.7 &  53.1 &  53.1 \\
RF n100 & 100 &    2.8 &  5.5 &  2.9 &    64.3 &  50.2 &  63.5 \\
       & 200 &    1.6 &  3.6 &  3.0 &    68.2 &  55.6 &  58.9 \\
       & 500 &    1.6 &  2.2 &  2.2 &    72.2 &  66.7 &  66.2 \\
RF n200 & 100 &    1.6 &  2.9 &  2.6 &    71.3 &  59.6 &  60.6 \\
       & 200 &    2.8 &  2.3 &  2.3 &    65.4 &  69.1 &  69.1 \\
       & 500 &    2.6 &  2.6 &  3.1 &    64.6 &  64.4 &  60.7 \\
RF n50 & 100 &    2.7 &  5.2 &  3.2 &    64.0 &  44.7 &  57.2 \\
       & 200 &    2.7 &  2.8 &  1.3 &    59.5 &  62.4 &  \textbf{73.7} \\
       & 500 &    2.1 &  4.2 &  2.6 &    63.8 &  48.3 &  60.2 \\
SVM c1 & 100 &    5.8 &  5.4 &  5.4 &    44.6 &  46.2 &  46.2 \\
       & 200 &    4.2 &  4.1 &  4.2 &    49.1 &  49.3 &  49.0 \\
       & 500 &    3.7 &  3.7 &  3.7 &    50.0 &  50.0 &  50.0 \\
\bottomrule
\end{tabular}
\end{table*}

\begin{figure}[hbt!]
  \centering
  \includegraphics[width=0.5\textwidth]{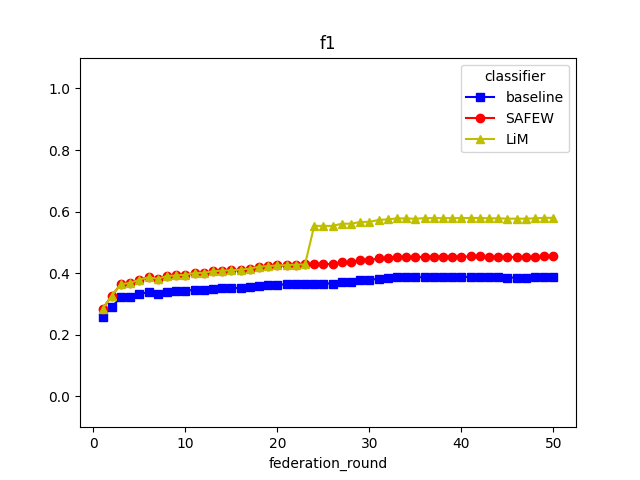}
  \caption{F1 score of 200 clients with 500 features, using kNN (n=3) as baseline classifier. SAFEW (i.e. no FL) improves on baseline, and LiM improves on SAFEW.\label{fig:client_f1_intervals}}
\end{figure}

Figure~\ref{fig:client_f1_intervals} shows the evolution of LiM using KNN as baseline and 500 features. Round 24 of the federation brings a significant improvement in performance for LiM, and then the F1 score slowly grows similarly to how SAFEW (i.e. no FL) and baseline do. By round 50, LiM reaches close to an F1 score of 70\%.


\subsubsection{Run-time performance}
\label{sec:runtime-performance}

The cloud and the clients compute their LiM predictions in the following three steps:
\begin{enumerate}
\item Build a local SAFEW. New weights are computed using testing data.
\item Average local and federated weights
\item Compute new predictions using the new weights
\end{enumerate}

For step 1, the cloud takes 13 seconds and an individual client, at the beginning of the federation, spends 0.1 seconds, due to the varying size of testing datasets. The cost of Step 2 is negligible. Step 3 however, takes almost the same time as step 1, i.e. step 1 spends only $1/10$ of the time computing new weights, which in step 3 is not necessary. Thus, clients spent 0.2 seconds computing the LiM final predictions, while the cloud spent 26 seconds.

It is worth noting that we did not apply any optimization to the current implementation of LiM, as our goal for this paper was to show that it is possible to build an FL-based malware detection while maintaining users' privacy. For additional performance gain -- at the very minimum, the predictions for base leaners from step 1 can be reused in step 3, making the total time closer to the time of step 1.


Regarding the training of the base learners, on average each model takes 12.7 seconds and the first cloud SAFEW prediction takes 13 seconds to complete. 

\subsection{Security analysis}
\label{subsec:security-analysis}

In section~\ref{sec:threat-model} we introduced two adversaries: one that aims to poison the federation process in order to make a specific malware app be misclassified as clean (Section~\ref{subsubsec:integrity-attacks}), and another that wants to learn which apps users have installed (Section~\ref{subsubsec:privacy-attacks}).

\subsubsection{Integrity attacks}
\label{subsubsec:integrity-attacks}

The considered adversary participates in the federation by controlling $50\%$ of clients, and attempts to bias the overall federation model by submitting maliciously crafted client models and malware apps. The end goal of the adversary is to trick LiM into misclassifying a malware app as clean. 

To make the malware app undetectable to LiM (i.e. a false negative), the adversary modifies the weights submitted by malicious clients so that his \texttt{allied base learners} have an honest majority in the different layers of the federation. A base learner is an ally of the adversary if it classifies the malware app as clean. We assume there is at least one ally in the LiM configuration for each round in the FL process.

We formalize the problem in equations~\ref{eq:adversary-goal} through~\ref{eq:cloud-constraint}.

Let $w$ be the honest weights of a malicious client, and $w'$ the poisoned weights. The adversary's goal is to make the two types of weights as similar as they can be in order for the attack to be stealthy. Equation~\ref{eq:adversary-goal} expresses this goal.

\begin{equation}
  \label{eq:adversary-goal}
  \begin{aligned}
    \underset{w'}{\text{minimize}}
    \norm{w - w'}
  \end{aligned}
\end{equation}

To maximize the chances of successfully poisoning the federation, the weights need to take into account multiple constraints. First, they must add up to one. Let $b$ be the number of base learners; then the first constraint to the optimization problem is expressed in equation~\ref{eq:add-up-to-one}.

\begin{equation}
  \label{eq:add-up-to-one}
  \begin{aligned}
    \sum_{i=1}^{b} w'_{i} = 1
  \end{aligned}
\end{equation}

Second, for the compromised client to misclassify the targeted app, the weights of the classifiers that err in favour of the adversary must account for an honest majority. Let $M$ be the indices of those allied classifiers in the SAFEW ensemble; then equation~\ref{eq:local-constraint} expresses the \texttt{local constraint}:

\begin{equation}
  \label{eq:local-constraint}
  \begin{aligned}
    \sum_{i=1}^{b} w'_{i} > 0.5, \{i \mid i \in M\}
  \end{aligned}
\end{equation}

Third, the cloud will average the weights of all the clients. Thus, the averaged weights of the allied classifiers must also hold an honest majority. Since the adversary does not have access to the weights of the honest clients, we can approximate it by averaging the \texttt{honest} weights of the malicious clients. Let $w^{c}$ be the weights of the client $c$; then equation~\ref{eq:clients-constraint} expresses the \texttt{clients constraint}.

\begin{equation}
  \label{eq:clients-constraint}
  \begin{aligned}
    \frac{1}{N+1}\sum_{c=1}^{N}\sum_{i=1}^{b}(w^{c}_{i} + w'_{i}) > 0.5, \{i \mid {i \in M} > 0.5\}
  \end{aligned}
\end{equation}

Finally, the cloud will compute the federated weights by averaging its own weights with the average of the clients. Assuming the weights of the cloud are known, we can make the same honest majority across allied classifiers hold by approximating it with the guessed average in equation~\ref{eq:clients-constraint}. Let $w^{*}$ be these averaged weights and $w^{cloud}$, the weights of the cloud; then equation~\ref{eq:cloud-constraint} expresses the \texttt{cloud constraint}.

\begin{equation}
  \label{eq:cloud-constraint}
  \begin{aligned}
    \frac{1}{2}\sum_{i=1}^{b}(w^{cloud}_{i} + w^{*}_{i}) > 0.5, \{i \mid {i \in M} > 0.5\},\\
  \end{aligned}
\end{equation}

Each round, the adversary has to try to solve this problem and submit the poisoned weights. If no solution is found, the problem is relaxed by first dropping the cloud constraint and then the client constraint. The adversary will always find a way to meet the local constraint.

All the malicious clients install the same app targeted by the adversary. We assume the adversary can partially modify the features of the app, which is crafted so that there is at least one allied base learner whose added weight(s) lie between $0.3$ and $0.4$. This range makes it possible for the adversary to succeed by redistributing up to 20\% of the honest weights, a relatively small percentage.

We evaluate the effectiveness of this attack on LiM for different configurations of learners both at the cloud and at the clients, with special focus on the latter as that is the final target of the adversary. Our results show that the poisoning attack has virtually no effect in the LiM cloud, with the F1 score plateauing at around 94\%--96\%, regardless of the configuration (cf. table~\ref{tab:poisoning-cloud} from the appendix).

Table~\ref{tab:poisoning-clients} compares the average client performance for honest LiM clients with baseline, SAFEW and adversarial (i.e. poisoned) clients. Overall, we see that the performance of LiM clients mostly outperforms baseline and SAFEW, and it is significantly higher than for adversarial clients. To better understand these results, we look at two specific configurations and analyze the effects of the attack over multiple federation rounds.

\begin{table*} [hbt!]
  \centering
  \caption{Comparison of client average performance across LiM, SAFEW and different baselines when 50\% of the clients are adversarial. Experiments are carried out for 50 rounds of federation using the Top 100, 200 and 500 features which were selected based on the chi2 test.\label{tab:poisoning-clients}}
  \begin{tabular}{l|c|cccc|cccc}
    \toprule
Baseline & \#Features & \multicolumn{4}{c|}{FP (raw values)} & \multicolumn{4}{c}{F1 (\%)} \\ \hline
       &  & Baseline &    SAFEW  &  LiM & Adv. client & Baseline &    SAFEW  &  LiM &  Adv. client \\
    \midrule
KNN n3 & 100 &   10.3 &  5.7 &  2.3 &    2.9 &    25.3 &  36.7 &  59.8 &    54.0 \\
       & 200 &    5.9 &  2.4 &  1.7 &    2.7 &    39.2 &  61.0 &  47.0 &    27.5 \\
       & 500 &    6.9 &  3.2 &  2.4 &    3.9 &    45.8 &  66.5 &  68.9 &    43.8 \\
LR c1 & 100 &    5.6 &  8.7 &  8.1 &    8.9 &    55.6 &  50.4 &  51.8 &    36.8 \\
       & 200 &    3.1 &  1.6 &  1.6 &    1.1 &    52.7 &  81.4 &  81.4 &    26.7 \\
       & 500 &    2.5 &  2.5 &  2.5 &    2.5 &    61.1 &  61.1 &  61.1 &    42.7 \\
RF n100 & 100 &    2.6 &  4.7 &  4.1 &    4.6 &    61.9 &  51.2 &  53.0 &    37.9 \\
       & 200 &    1.1 &  1.6 &  1.0 &    1.6 &    70.0 &  75.1 &  78.9 &    50.5 \\
       & 500 &    2.2 &  2.2 &  2.2 &    2.1 &    64.1 &  64.1 &  64.2 &    53.8 \\
RF n200 & 100 &    2.5 &  3.0 &  3.0 &    3.0 &    55.0 &  50.0 &  50.0 &    16.3 \\
       & 200 &    1.5 &  2.6 &  2.1 &    3.0 &    66.5 &  64.5 &  65.7 &    39.9 \\
       & 500 &    2.2 &  2.5 &  2.2 &    4.6 &    56.8 &  54.8 &  56.8 &    18.1 \\
RF n50 & 100 &    1.8 &  2.8 &  2.5 &    3.3 &    72.5 &  64.2 &  65.3 &    54.7 \\
       & 200 &    2.1 &  4.2 &  1.8 &    2.2 &    67.6 &  61.6 &  73.2 &    43.9 \\
       & 500 &    3.1 &  2.5 &  2.5 &    2.6 &    58.3 &  62.3 &  62.3 &    36.9 \\
SVM c1 & 100 &    5.1 &  5.1 &  5.1 &    4.0 &    41.5 &  41.5 &  41.5 &    20.5 \\
       & 200 &    4.0 &  3.5 &  3.5 &    0.5 &    49.1 &  52.7 &  52.7 &    55.5 \\
       & 500 &    3.7 &  3.7 &  3.7 &    2.6 &    42.7 &  42.7 &  42.7 &    12.1 \\
    \bottomrule
  \end{tabular}
\end{table*}

\begin{figure}  [h]
  \centering
  \includegraphics[width=0.5\textwidth]{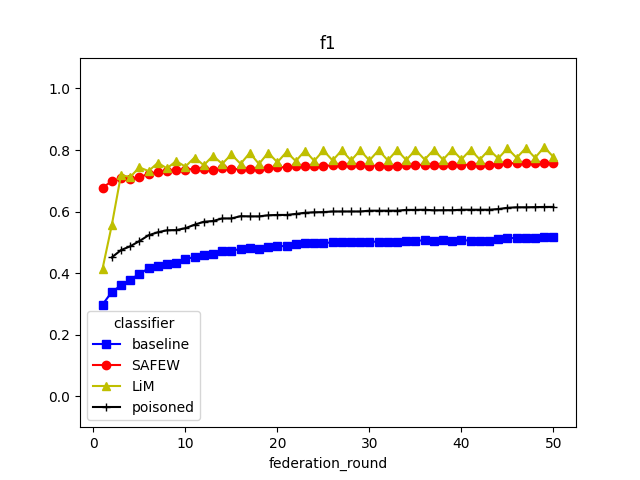}
  \caption{F1 score of 100 honest + 100 malicious clients, using 200 features with baseline classifier kNN (n=3). LiM outperforms SAFEW  (i.e. no FL), which performs better than baseline.\label{fig:client_f1_intervals_poisoned}}
\end{figure}

\begin{figure}  [h]
  \centering
  \includegraphics[width=0.5\textwidth]{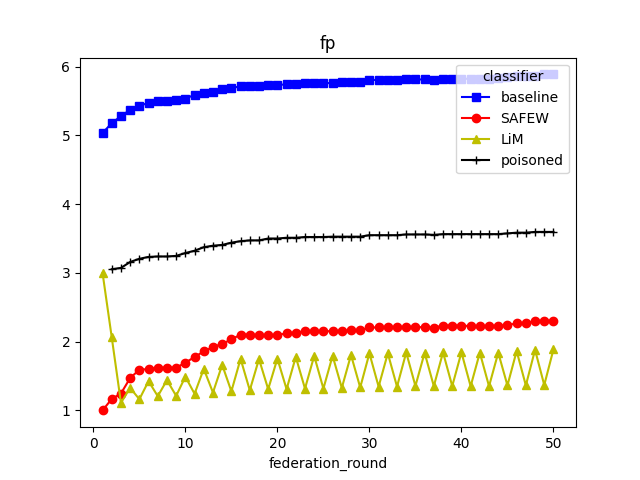}
  \caption{False positives of 200 clients with 200 features, using kNN (n=3) as  baseline classifier. \label{fig:client_fp_intervals_poisoned}}
\end{figure}

We first consider kNN with $n=3$ and 200 features as baseline classifier. Figure~\ref{fig:client_f1_intervals_poisoned} shows that LiM clients perform better than baseline and SAFEW. We can also see that client performance improves over time, with the most significant gains taking place in the first federation rounds. The figure also shows how the F1 score of the adversarial clients approximates that of SAFEW, suggesting that the adversary is indeed submitting poisoned weights that are not too different from honest weights. The actual number of false positives for this configuration is shown in figure~\ref{fig:client_fp_intervals_poisoned}. As can be seen in the figure, LiM clients have a small number of false positives compared to baseline, SAFEW and poisoned clients. Note that the adversary's goal is to trigger a false negative, and thus the number of false positives in adversarial clients can be minimized to behave as close to an honest client as possible. Due to the small number of malware apps installed by clients, a small increase in the number of false positives has a big impact on the F1 score.

\begin{figure} [h]
  \centering
  \includegraphics[width=0.5\textwidth]{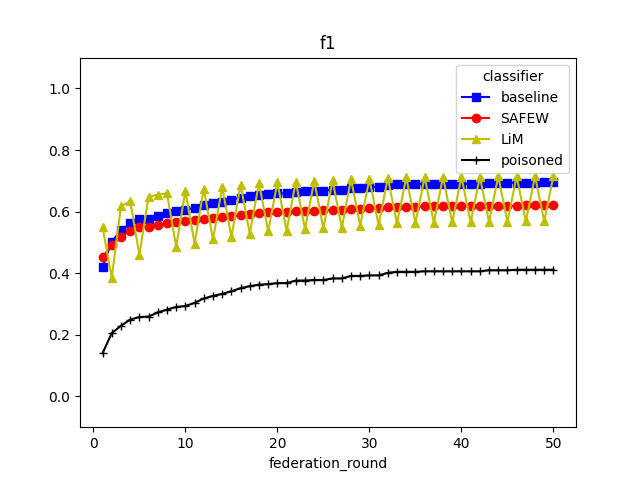}
  \caption{F1 score of 200 clients with 200 features, using RFn200 as baseline classifier and with poisoning. \label{fig:client_f1_svm_poisoned}}
\end{figure}

\begin{figure}   [h]
  \centering
  \includegraphics[width=0.5\textwidth]{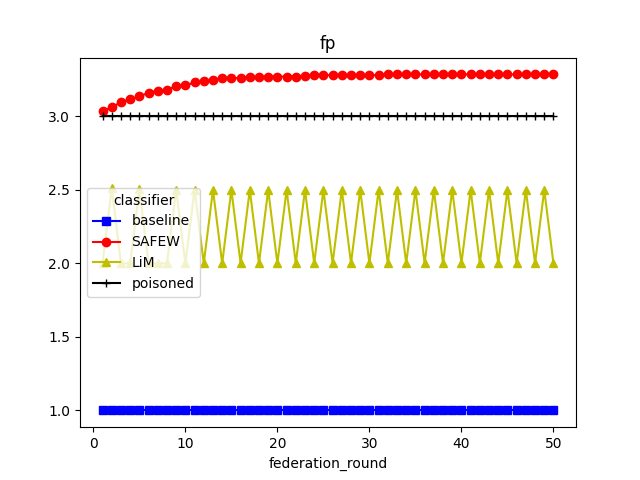}
  \caption{False positives of 200 clients with 200 features, using RFn200 as baseline classifier and with poisoning. \label{fig:client_fp_svm_poisoned}}
\end{figure}

For the second case, we consider as baseline classifier an RF with 200 decision trees and $200$ features. In the results shown in figure~\ref{fig:client_f1_svm_poisoned}, we can see that clients are affected by the attack every second round. The F1 score jumps between 0.4 and 0.6 due to slight variations of the weights that make the false positives per client jump from 2 to 2.5, as shown in figure~\ref{fig:client_fp_svm_poisoned}.

These results indicate that LiM is resistant against poisoning attacks.
This is thanks to the strong influence of the cloud in the federated weights and the use of trusted data (cf.~\cite{shejwalkar_manipulating_2021} - L2 and L3 from section 7-A) to compute the cloud weights. Moreover, the power of the attacker is limited by the small number of SAFEW weights, which makes it difficult to distribute noise across multiple dimensions (cf.~\cite{shejwalkar_manipulating_2021} - L1 from section 7-A) This allows LiM to use malicious and honest weights together and still resist poisoning attacks.

\subsubsection{Privacy attacks}
\label{subsubsec:privacy-attacks}

Federated learning provides a defense mechanism against privacy attacks: hiding the raw features of the installed apps to make it more difficult for the cloud server to infer information. However, the submitted client models may still leak information that can be used to infer, e.g. if a specific app has been part of their training set. Membership inference~\cite{melis_exploiting_2019} relies on the fact that updates to the client models may change only a few parameters, and those parameters can reveal information about specific apps used locally in a federation round.

In LiM, updates to client models only change the SAFEW weights of the base learners. It is safe to assume that the number of base learners of SAFEW is significantly smaller than the number of parameters of an individual base learner, e.g. a deep neural network. This data compression greatly reduces the information available to the adversary to perform the membership inference, lowering the probability of success of this attack.

We evaluate if the attack by Melis et al.~\cite{melis_exploiting_2019} can infer whether clients installed a target app, using the weights shared by the clients as the gradients. We denote by $r$ the federation round, $i$ is the $i-$th client, and $c$ the cloud. The cloud first computes the weights associated with a specific app,
$w_{app}^{r} = SAFEW(baseline\_model, base\_models, app)$, and the weights of the client with no federation for round $r$, undoing step 5 as described in section~\ref{sec:architecture}:
$w_{c}^{r, no federated} = (2 * w_{i}^{r}) - w_{c}^{r-1}$

Clients compute the weights $w_{c}$ using the data from all the apps they have installed. We assume the cloud does not have access to the list of preinstalled apps of each client, but aims to identify the apps users install in each round. To achieve that goal, the cloud isolates the contribution of a single round by subtracting the weights of the last and the previous rounds, i.e. round $r$ and $r-1$. This is reflected in equation~\ref{eq:membership-inference}. The attack succeeds if equation~\ref{eq:membership-inference} holds and the identified app was indeed installed by the client in the previous round (i.e. there is a true positive).

\begin{equation}
  \label{eq:membership-inference}
  w_{app}^{r} - (w_{c}^{r, no federated} - w_{c}^{r-1, no federated}) = 0
  \end{equation}

On the contrary, if there is no app that makes equation~\ref{eq:membership-inference} hold, then the attack does not succeed. If equation~\ref{eq:membership-inference} is satisfied by 2 or more apps that leads to false positives. In that case, the best the adversary can do is to guess among the candidate apps, so the more apps that produce the same weights, the more resistant is LiM to privacy attacks.

We implemented this attack and ran experiments with the same SAFEW configurations (sets of baseline and base learners). In order to increase the likelihood of success, we let the cloud have access to the complete dataset of apps: this way, clients cannot install apps that the cloud cannot infer membership of. For each round, the cloud counts the number of apps in its dataset that makes equation~\ref{eq:membership-inference} hold. We then check if those apps were indeed installed by the corresponding client.

The results show that the cloud is unable to find any app that makes equation~\ref{eq:membership-inference} hold, i.e. there are 0 true positives and 0 false positives. We observe that the weights resulting from testing a SAFEW model with a single specific app are very different from the difference between the weights sent in the last two rounds. Specifically, we realize that there is not enough information in a single app for SAFEW to associate different weights to each base learner, while in successive federation rounds, clients slightly increase some weights in detriment of others.

\subsection{Evaluation using MaMaDroid dataset}

In order to further verify the performance of LiM, we also evaluated LiM using the MaMaDroid dataset~\cite{onwuzurike_mamadroid:_2019} under the setup described at the beginning of this section. Due to errors in the download and the feature extraction, we used 7K malware and 7K clean apps. We ran 2 experiments per configuration without adversarial clients, and 2 experiments where half of the clients perform the poisoning attack described in section~\ref{subsubsec:integrity-attacks}.

First, we will describe the results of the experiments without adversarial clients. Then, we will focus on the results obtained with 50\% of clients submitting poisoned weights. For each of the scenarios, we analyze the cloud and the client results.

\subsubsection{Without poisoning} 

We observed no performance difference between LiM, SAFEW, Centralized SAFEW, and baseline in the cloud, as they all obtain an F1 score of around 90\%. On the other hand, there was a noticeable difference in the performance of the clients.
Similarly to the results presented in previous sections, LiM is able to outperform baseline and SAFEW most of the times, 
including in configurations that use RF (n=50,200) as baseline.

Overall, the F1 scores were lower with this new dataset due to the higher number of false positives in the baseline learners. This is not surprising, since the dataset itself has fewer distinguishable manifest features and even important dynamic analysis features are present in both clean and malicious apps, resulting in poor distinguishers (cf. section 4.4 from~\cite{onwuzurike_mamadroid:_2019}).

\subsubsection{With poisoning} 

More interesting were the results of experiments where 50\% of the clients poisoned their weights to trigger a false negative, i.e. so that a specific malicious app is misclassified as clean. Table~\ref{tab:poisoning-clients-mamadroid} shows that LiM still outperforms baseline and SAFEW in most of the cases. The highest performance gain with respect to the baseline and SAFEW is using kNN with 100 features as baseline.

\begin{figure}  [h]
  \centering
  \includegraphics[width=0.5\textwidth]{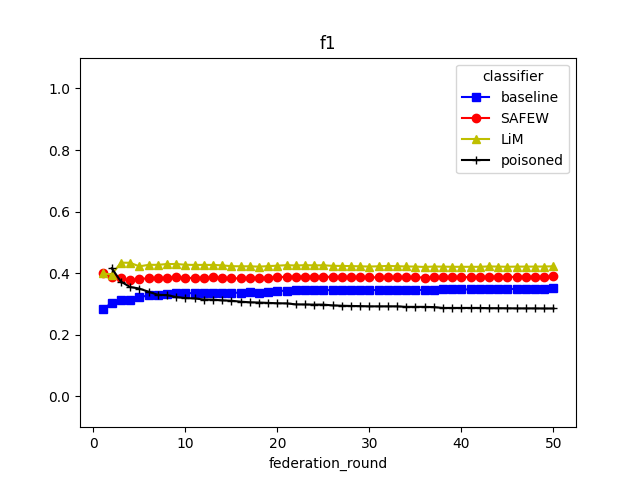}
  \caption{F1 score of 200 clients and the MaMaDroid dataset with 500 features, using kNN (n=3) as baseline classifier~\label{fig:client_f1_mamadroid_500_features_baseline_knn3_poisoned}}
\end{figure}

An example of such configuration is shown in figure~\ref{fig:client_f1_mamadroid_500_features_baseline_knn3_poisoned}. We can see that LiM outperforms both SAFEW and baseline, stabilizing at $0.4$ even as the poisoned clients worsen their performance. This shows that LiM, under a targeted integrity attack, can still learn from the federation.

\begin{table*} [hbt!]
\centering
\caption{Comparison of client average performance using the MaMaDroid dataset across LiM, SAFEW and different baselines when 50\% of the clients are adversarial. Experiments are carried out for 50 rounds of federation using the Top 100, 200 and 500 features which were selected based on the chi2 test.\label{tab:poisoning-clients-mamadroid}}
\begin{tabular}{l|c|cccc|cccc}
  \toprule
Baseline & \#Features & \multicolumn{4}{c|}{FP (raw values)} & \multicolumn{4}{c}{F1 (\%)} \\ \hline
       &  & Baseline & SAFEW & LiM & Adv. client & Baseline  &  SAFEW & LiM & Adv. client \\
\midrule
KNN n3 & 100 &    8.5 &   8.5 &   5.7 &    5.4 &    25.1 &  23.8 &   32.5 &    24.3 \\
       & 200 &    7.3 &   5.9 &   5.2 &    5.2 &    32.5 &  41.3 &   39.5 &    21.7 \\
       & 500 &    6.1 &   5 &   4.6 &    4.5 &    34.4 &  38.7 &   40.6 &    30.7 \\
LR c1 & 100 &    7.8 &   7.5 &   7.1 &    7.1 &    26.4 &  27.2 &   28.8 &    11.1 \\
       & 200 &    8.4 &   8.6 &   8.6 &    7.5 &    20.4 &  15.2 &   17.5 &    0 \\
       & 500 &    5.3 &   5.0 &   4.1 &    3.6 &    17.8 &  33.0 &   37.5 &    39.8 \\
RF n100 & 100 &    6.7 &   8.9 &   6.8 &    6.8 &    26.4 &  26.9 &   32.9 &    15.9 \\
       & 200 &    8.3 &  10.8 &   9.5 &    9.3 &    44.2 &  38.4 &   40.2 &    29.8 \\
       & 500 &    5.4 &   5.4 &   5.3 &    4.3 &    36.7 &  36.7 &   36.6 &    22.0 \\
RF n200 & 100 &    7.3 &  10.1 &   7.8 &    6.9 &    42.0 &  34.2 &   39.8 &    39.7 \\
       & 200 &    5.1 &   5.1 &   5.1 &    5 &    28.5 &  28.5 &   28.5 &    0.9 \\
       & 500 &    4.3 &   4.1 &   3.2 &    3.8 &    38.5 &  39.1 &   46.5 &    25.8 \\
RF n50 & 100 &    6.4 &   9.9 &   7.0 &    5.9 &    29.1 &  19.8 &   22.5 &    0.3 \\
       & 200 &    5.2 &   6.5 &   5.2 &    4.9 &    40.1 &  35.4 &   41.0 &    49.1 \\
       & 500 &    5.9 &   6.5 &   5.4 &    5.0 &    42.6 &  41.0 &   44.6 &    43.8 \\
SVM c1 & 100 &    7.2 &   7.7 &   7.3 &    7.2 &    23.2 &  24.0 &   23.3 &    10.7 \\
       & 200 &   10.2 &   7.6 &   6.2 &    5.2 &    20.2 &  25.9 &   28.5 &    25.6 \\
       & 500 &    2 &   4 &   4 &    5.2 &    60.9 &  41.8 &   41.8 &    35.1 \\
\bottomrule
\end{tabular}
\end{table*}

Finally, we found the membership inference attack unsuccessful, as the cloud is unable to find any single app that makes equation~\ref{eq:membership-inference} hold in any of these experiments (0 true positives, 0 false positives), further indicating that LiM is resistant to membership inference.

\section{Discussion and future work}\label{sec:discussion}

Our evaluation shows that LiM enables federated clients to learn from each other even when they do not have local access to ground truth. Our experiments show the importance of the learner selection in the individual SAFEWs, which must maximize variance across their predictions as well as their individual performance. We observe that random forests provide a high F1 score by themselves, but LiM can outperform them by fine-tuning the weights of weaker but complementary learners. The best results were achieved when all learners influenced the final predictions, rather than defaulting to the predictions of the strongest learner. In this paper we have used standard learners, but we expect LiM to greatly benefit from a more careful selection of base learners. Future work is needed to verify this hypothesis.

Further measures to protect users from privacy attacks could include differential privacy. Previous work has shown that adding noise to the client parameters can prevent information leakage from client models~\cite{wei_federated_2020}. LiM takes instead a data aggregation approach, compressing client parameters to a smaller set of bounded hyper-parameters to significantly decrease the amount of information available to the adversary (cf. section~\ref{subsec:security-analysis}). This approach does not compromise on the quality of the models shared by the users, and it does not introduce limitations regarding the minimum number of users in the federation~\cite{wei_federated_2020}. Nevertheless, LiM's protection does not hide whether or not users have installed new applications, as users' models remain unchanged while no new applications are installed; differential privacy can stop the leakage of such information.

Regarding performance results, we highlight that 1) LiM does not sacrifice performance for privacy, as it matches and sometimes even outperforms a privacy-invasive Centralized SAFEW, and 2) the number of false positives in clients decreases across federation rounds. We observe a drastic improvement in the first round of federation, when client weights contribute to updating the initial model created by the cloud. This leads us to believe that increasing the number of clients may reproduce this effect along further rounds, as there will be more information coming from additional client weights. Simulating LiM at scale can help clarify the relationship between the number of clients and LiM performance.

Interestingly, LiM can avoid virtually any privacy loss with respect to a centralized SAFEW as the federation of the clients provides enough information to arrive to the same weights of the privacy-invasive model. Even though the differences between the weights of the cloud without federation and the weights of the centralized SAFEW can be relatively high (e.g. to 0.1 in a single weight, 0.37 vs 0.47), clients provide enough information to equalize them.

We envision LiM as a system that can be practically deployed in real-world smartphones. While market interests may dissuade powerful organizations like Google to deploy LiM in stock Android, we believe third-party Android distributions differentiating themselves by being more privacy conscious can develop the client app as a privileged service executed upon the installation of (one or more) apps. This practical implementation would perform static analysis over the manifest files of the newly installed apps, without the need to trust the LiM service provider with sensitive client models.

\textbf{Limitations \& Future work:} We expect future work to address the following limitations of the current formulation of LiM, namely:
\begin{itemize}
\item Malware family-wise classification is out of the scope in this paper. Our figures do not take into account the specific characteristics of the malware apps installed by clients.
\item Real world experiments at scale. In this paper, we perform experiments with standard learners, 50 rounds of federation, a static set of users, and pre-defined parameters to simulate probabilities of installing malware, clean and popular apps. Increasing the number of rounds and performing a thorough selection of the SAFEW learners would provide a more thorough understanding of the performance of LiM in different practical scenarios.
\end{itemize}

We believe LiM is equipped to act as a self-evolving system that can update itself using the ever-enlarging set of apps users install on their devices. This low maintenance overhead of the service provider, which does not need to label all data to make use of it, together with the geographic distribution of malware, can help LiM detect malicious apps faster~\cite{hutchinson_are_2019} and thus prevent malware to disseminate in large numbers. Other applications where self-updating models are of interest could explore LiM as a solution where no burden is placed to the user nor the service provider.

\section{Related work}
\label{sec:relatedwork}

\subsection{Machine learning for Android malware classification}
ML techniques to detect mobile malware have been extensively investigated, leveraging a few characteristics of the mobile applications (for example, call graphs~\cite{mirzaei2019andrensemble}, permissions~\cite{li2018significant}, or both API calls and permissions~\cite{kim2018multimodal}), and the results obtained were promising. Classification approaches have also been proposed to model and approximate the behaviors of Android applications and discern malicious apps from benign ones. The detection accuracy of a classification method depends on the quality of the features (for example, how specific the features are~\cite{MoonsamyMiningpermissionpatterns2014}).

In~\cite{MilosevicMachinelearningaided2017}, Milosevic et al. implemented an app that detects malware locally through a pre-trained SVM classifier. They explicitly created a permission based model to detect malware. There is no information to recreate the model, although the model itself is available as part of the OWASP Seraphimdroid project~\cite{OWASPSeraphimDroidProject}. The dataset they used has 200 benign and 200 malicious apps; however, since 2015, their dataset is no longer publicly available. For more recent state of the art related work in this area, we refer the reader to~\cite{onwuzurike_mamadroid:_2019}.



\subsection{Semi-supervised federated learning}


Semi-supervised FL has already been proposed by~\cite{albaseer_exploiting_2020} to take advantage of the abundant unlabeled data in the smart city context. Contrary to LiM, they assume a subset of clients have labeled samples, and use them to train a classifier that will provide the missing labels to retrain another local model. They report an improvement of 8\% accuracy compared to the fully supervised FL, but do not consider privacy and integrity attacks nor provide a lower bound for the performance of the clients and the server.








\section{Conclusion}
\label{sec:conclusion}

We have presented LiM, the first FL-based malware detection framework that works successfully without user access to ground truth by making use of safe semi-supervised learning techniques. We demonstrate its utility as a hybrid system where users keep their apps secret from the service provider while successfully detecting most of the malicious apps they install without raising many false alarms.
LiM is resistant against a strategic adversary that compromises 50\% of the clients and crafts a malicious app in order to bypass the detection mechanism. Our proposed tool can also withstand membership inference attacks that exploit client updates to try to determine if a specific app was installed by a client. 
While we have evaluated LiM in the malware detection domain, we note that it can be applied to other problems where users cannot provide ground truth labels to their clients' models, but still benefit from FL -- both in terms of performance improvements and privacy properties.

\section*{Acknowledgements}
This research is partially supported by BMK, BMDW, and the Province of Upper Austria in the frame of the COMET Programme managed by FFG in the COMET Module S3AI, by the Research Council KU Leuven under the grant C24/18/049, by the Deutsche Forschungsgemeinschaft (DFG, German Research Foundation) under Germany's Excellence Strategy - EXC 2092 CaSa - 390781972, by CyberSecurity Research Flanders with reference number VR20192203, and by the United States Air Force and DARPA under Contract No. FA8750-19-C-0502. Any opinions, findings and conclusions or recommendations expressed in this material are those of the authors and do not necessarily reflect the views of the United States Air Force and DARPA. 

We also thank Brecht Van der Vliet for his helpful review of the code and the identification of an important implementation issue.

\bibliographystyle{plain}
\bibliography{main}

\appendix

\section{List of features}
\subsection{Top 100 features}
\tiny
\begin{itemize}
\item android\_hardware\_camera
\item android\_hardware\_camera\_autofocus
\item android\_hardware\_microphone
\item android\_hardware\_screen\_landscape
\item android\_hardware\_screen\_portrait
\item android\_hardware\_touchscreen\_multitouch
\item android\_hardware\_touchscreen\_multitouch\_distinct
\item android\_hardware\_wifi
\item android\_permission\_access\_assisted\_gps
\item android\_permission\_access\_coarse\_location
\item android\_permission\_access\_coarse\_updates
\item android\_permission\_access\_fine\_location
\item android\_permission\_access\_gps
\item android\_permission\_access\_location
\item android\_permission\_access\_location\_extra\_commands
\item android\_permission\_access\_wifi\_state
\item android\_permission\_call\_phone
\item android\_permission\_change\_wifi\_state
\item android\_permission\_get\_tasks
\item android\_permission\_install\_packages
\item android\_permission\_kill\_background\_processes
\item android\_permission\_mount\_unmount\_filesystems
\item android\_permission\_process\_outgoing\_calls
\item android\_permission\_read\_call\_log
\item android\_permission\_read\_contacts
\item android\_permission\_read\_logs
\item android\_permission\_read\_phone\_state
\item android\_permission\_read\_profile
\item android\_permission\_read\_settings
\item android\_permission\_read\_sms
\item android\_permission\_receive\_boot\_completed
\item android\_permission\_receive\_sms
\item android\_permission\_restart\_packages
\item android\_permission\_send\_sms
\item android\_permission\_system\_alert\_window
\item android\_permission\_use\_credentials
\item android\_permission\_write\_apn\_settings
\item android\_permission\_write\_contacts
\item android\_permission\_write\_external\_storage
\item android\_permission\_write\_settings
\item android\_permission\_write\_sms
\item android\_support\_v4\_content\_fileprovider
\item cn\_domob\_android\_ads\_domobactivity
\item com\_adfeiwo\_ad\_coverscreen\_sa
\item com\_adfeiwo\_ad\_coverscreen\_sr
\item com\_adfeiwo\_ad\_coverscreen\_wa
\item com\_adwo\_adsdk\_adwoadbrowseractivity
\item com\_airpush\_android\_deliveryreceiver
\item com\_airpush\_android\_messagereceiver
\item com\_airpush\_android\_pushads
\item com\_airpush\_android\_pushservice
\item com\_airpush\_android\_userdetailsreceiver
\item com\_android\_browser\_permission\_read\_history\_bookmarks
\item com\_android\_browser\_permission\_write\_history\_bookmarks
\item com\_android\_launcher\_permission\_install\_shortcut
\item com\_android\_launcher\_permission\_uninstall\_shortcut
\item com\_android\_vending\_billing
\item com\_bving\_img\_ag
\item com\_bving\_img\_rv
\item com\_bving\_img\_se
\item com\_facebook\_ads\_interstitialadactivity
\item com\_facebook\_facebookactivity
\item com\_facebook\_loginactivity
\item com\_google\_android\_c2dm\_permission\_receive
\item com\_google\_android\_gms\_ads\_adactivity
\item com\_google\_android\_gms\_ads\_purchase\_inapppurchaseactivity
\item com\_google\_android\_gms\_analytics\_analyticsreceiver
\item com\_google\_android\_gms\_analytics\_analyticsservice
\item com\_google\_android\_gms\_analytics\_campaigntrackingreceiver
\item com\_google\_android\_gms\_analytics\_campaigntrackingservice
\item com\_google\_android\_gms\_appinvite\_previewactivity
\item com\_google\_android\_gms\_auth\_api\_signin\_internal\_signinhubactivity
\item com\_google\_android\_gms\_auth\_api\_signin\_revocationboundservice
\item com\_google\_android\_gms\_common\_api\_googleapiactivity
\item com\_google\_android\_gms\_gcm\_gcmreceiver
\item com\_google\_android\_gms\_measurement\_appmeasurementcontentprovider
\item com\_google\_android\_gms\_measurement\_appmeasurementinstallreferrerreceiver
\item com\_google\_android\_gms\_measurement\_appmeasurementreceiver
\item com\_google\_android\_gms\_measurement\_appmeasurementservice
\item com\_google\_android\_providers\_gsf\_permission\_read\_gservices
\item com\_google\_firebase\_iid\_firebaseinstanceidinternalreceiver
\item com\_google\_firebase\_iid\_firebaseinstanceidreceiver
\item com\_google\_firebase\_iid\_firebaseinstanceidservice
\item com\_google\_firebase\_messaging\_firebasemessagingservice
\item com\_google\_firebase\_provider\_firebaseinitprovider
\item com\_google\_update\_dialog
\item com\_google\_update\_receiver
\item com\_google\_update\_updateservice
\item com\_kuguo\_ad\_boutiqueactivity
\item com\_kuguo\_ad\_mainactivity
\item com\_kuguo\_ad\_mainreceiver
\item com\_kuguo\_ad\_mainservice
\item com\_mobclix\_android\_sdk\_mobclixbrowseractivity
\item com\_soft\_android\_appinstaller\_\_finishactivity
\item com\_soft\_android\_appinstaller\_\_firstactivity
\item com\_soft\_android\_appinstaller\_\_rulesactivity
\item com\_soft\_android\_appinstaller\_memberactivity
\item com\_soft\_android\_appinstaller\_questionactivity
\item com\_startapp\_android\_publish\_appwallactivity
\item net\_youmi\_android\_adactivity
\end{itemize}

\subsection{Top 200 features}
\begin{itemize}
\item android\_hardware\_camera
\item android\_hardware\_camera\_autofocus
\item android\_hardware\_camera\_front
\item android\_hardware\_microphone
\item android\_hardware\_screen\_landscape
\item android\_hardware\_screen\_portrait
\item android\_hardware\_touchscreen\_multitouch
\item android\_hardware\_touchscreen\_multitouch\_distinct
\item android\_hardware\_wifi
\item android\_permission\_access\_assisted\_gps
\item android\_permission\_access\_coarse\_location
\item android\_permission\_access\_coarse\_updates
\item android\_permission\_access\_fine\_location
\item android\_permission\_access\_gps
\item android\_permission\_access\_location
\item android\_permission\_access\_location\_extra\_commands
\item android\_permission\_access\_wifi\_state
\item android\_permission\_call\_phone
\item android\_permission\_camera
\item android\_permission\_change\_configuration
\item android\_permission\_change\_network\_state
\item android\_permission\_change\_wifi\_state
\item android\_permission\_clear\_app\_cache
\item android\_permission\_get\_tasks
\item android\_permission\_install\_packages
\item android\_permission\_kill\_background\_processes
\item android\_permission\_mount\_unmount\_filesystems
\item android\_permission\_process\_outgoing\_calls
\item android\_permission\_read\_calendar
\item android\_permission\_read\_call\_log
\item android\_permission\_read\_contacts
\item android\_permission\_read\_external\_storage
\item android\_permission\_read\_logs
\item android\_permission\_read\_phone\_state
\item android\_permission\_read\_profile
\item android\_permission\_read\_settings
\item android\_permission\_read\_sms
\item android\_permission\_receive\_boot\_completed
\item android\_permission\_receive\_sms
\item android\_permission\_receive\_wap\_push
\item android\_permission\_record\_audio
\item android\_permission\_restart\_packages
\item android\_permission\_send\_sms
\item android\_permission\_system\_alert\_window
\item android\_permission\_use\_credentials
\item android\_permission\_vibrate
\item android\_permission\_write\_apn\_settings
\item android\_permission\_write\_calendar
\item android\_permission\_write\_contacts
\item android\_permission\_write\_external\_storage
\item android\_permission\_write\_settings
\item android\_permission\_write\_sms
\item android\_support\_v4\_content\_fileprovider
\item biz\_neoline\_android\_reader\_bookmarksandtocactivity
\item biz\_neoline\_android\_reader\_libraryactivity
\item biz\_neoline\_android\_reader\_neobookreader
\item biz\_neoline\_android\_reader\_textsearchactivity
\item biz\_neoline\_app\_core\_core\_application\_shutdownreceiver
\item biz\_neoline\_app\_core\_ui\_android\_dialogs\_dialogactivity
\item biz\_neoline\_app\_core\_ui\_android\_library\_crashreportingactivity
\item biz\_neoline\_test\_donationactivity
\item cn\_domob\_android\_ads\_domobactivity
\item com\_adfeiwo\_ad\_coverscreen\_sa
\item com\_adfeiwo\_ad\_coverscreen\_sr
\item com\_adfeiwo\_ad\_coverscreen\_wa
\item com\_adwo\_adsdk\_adwoadbrowseractivity
\item com\_adwo\_adsdk\_adwosplashadactivity
\item com\_airpush\_android\_deliveryreceiver
\item com\_airpush\_android\_messagereceiver
\item com\_airpush\_android\_pushads
\item com\_airpush\_android\_pushservice
\item com\_airpush\_android\_smartwallactivity
\item com\_airpush\_android\_userdetailsreceiver
\item com\_amazon\_device\_messaging\_permission\_receive
\item com\_anddoes\_launcher\_permission\_update\_count
\item com\_android\_browser\_permission\_read\_history\_bookmarks
\item com\_android\_browser\_permission\_write\_history\_bookmarks
\item com\_android\_launcher\_permission\_install\_shortcut
\item com\_android\_launcher\_permission\_uninstall\_shortcut
\item com\_android\_vending\_billing
\item com\_biznessapps\_layout\_maincontroller
\item com\_biznessapps\_player\_playerservice
\item com\_biznessapps\_pushnotifications\_c2dmmessagesreceiver
\item com\_biznessapps\_pushnotifications\_c2dmregistrationreceiver
\item com\_bving\_img\_ag
\item com\_bving\_img\_rv
\item com\_bving\_img\_se
\item com\_chartboost\_sdk\_cbimpressionactivity
\item com\_elm\_lma
\item com\_elm\_lmr
\item com\_elm\_lms
\item com\_elm\_lmsk
\item com\_facebook\_ads\_audiencenetworkactivity
\item com\_facebook\_ads\_interstitialadactivity
\item com\_facebook\_customtabactivity
\item com\_facebook\_customtabmainactivity
\item com\_facebook\_facebookactivity
\item com\_facebook\_facebookcontentprovider
\item com\_facebook\_loginactivity
\item com\_feiwothree\_coverscreen\_sa
\item com\_feiwothree\_coverscreen\_sr
\item com\_feiwothree\_coverscreen\_wa
\item com\_google\_android\_apps\_analytics\_analyticsreceiver
\item com\_google\_android\_c2dm\_permission\_receive
\item com\_google\_android\_gcm\_gcmbroadcastreceiver
\item com\_google\_android\_gms\_ads\_adactivity
\item com\_google\_android\_gms\_ads\_purchase\_inapppurchaseactivity
\item com\_google\_android\_gms\_analytics\_analyticsreceiver
\item com\_google\_android\_gms\_analytics\_analyticsservice
\item com\_google\_android\_gms\_analytics\_campaigntrackingreceiver
\item com\_google\_android\_gms\_analytics\_campaigntrackingservice
\item com\_google\_android\_gms\_appinvite\_previewactivity
\item com\_google\_android\_gms\_auth\_api\_signin\_internal\_signinhubactivity
\item com\_google\_android\_gms\_auth\_api\_signin\_revocationboundservice
\item com\_google\_android\_gms\_common\_api\_googleapiactivity
\item com\_google\_android\_gms\_gcm\_gcmreceiver
\item com\_google\_android\_gms\_measurement\_appmeasurementcontentprovider
\item com\_google\_android\_gms\_measurement\_appmeasurementinstallreferrerreceiver
\item com\_google\_android\_gms\_measurement\_appmeasurementjobservice
\item com\_google\_android\_gms\_measurement\_appmeasurementreceiver
\item com\_google\_android\_gms\_measurement\_appmeasurementservice
\item com\_google\_android\_providers\_gsf\_permission\_read\_gservices
\item com\_google\_firebase\_iid\_firebaseinstanceidinternalreceiver
\item com\_google\_firebase\_iid\_firebaseinstanceidreceiver
\item com\_google\_firebase\_iid\_firebaseinstanceidservice
\item com\_google\_firebase\_messaging\_firebasemessagingservice
\item com\_google\_firebase\_provider\_firebaseinitprovider
\item com\_google\_update\_dialog
\item com\_google\_update\_receiver
\item com\_google\_update\_updateservice
\item com\_htc\_launcher\_permission\_update\_shortcut
\item com\_klpcjg\_wyxjvs102320\_browseractivity
\item com\_klpcjg\_wyxjvs102320\_mainactivity
\item com\_klpcjg\_wyxjvs102320\_vdactivity
\item com\_kuguo\_ad\_boutiqueactivity
\item com\_kuguo\_ad\_mainactivity
\item com\_kuguo\_ad\_mainreceiver
\item com\_kuguo\_ad\_mainservice
\item com\_majeur\_launcher\_permission\_update\_badge
\item com\_mobclix\_android\_sdk\_mobclixbrowseractivity
\item com\_nd\_dianjin\_activity\_offerappactivity
\item com\_onesignal\_gcmbroadcastreceiver
\item com\_onesignal\_gcmintentservice
\item com\_onesignal\_notificationopenedreceiver
\item com\_onesignal\_permissionsactivity
\item com\_onesignal\_syncservice
\item com\_parse\_gcmbroadcastreceiver
\item com\_parse\_parsebroadcastreceiver
\item com\_parse\_pushservice
\item com\_paypal\_android\_sdk\_payments\_futurepaymentconsentactivity
\item com\_paypal\_android\_sdk\_payments\_futurepaymentinfoactivity
\item com\_paypal\_android\_sdk\_payments\_loginactivity
\item com\_paypal\_android\_sdk\_payments\_paymentactivity
\item com\_paypal\_android\_sdk\_payments\_paymentconfirmactivity
\item com\_paypal\_android\_sdk\_payments\_paymentmethodactivity
\item com\_paypal\_android\_sdk\_payments\_paypalfuturepaymentactivity
\item com\_paypal\_android\_sdk\_payments\_paypalservice
\item com\_sec\_android\_provider\_badge\_permission\_read
\item com\_sec\_android\_provider\_badge\_permission\_write
\item com\_soft\_android\_appinstaller\_\_finishactivity
\item com\_soft\_android\_appinstaller\_\_firstactivity
\item com\_soft\_android\_appinstaller\_\_rulesactivity
\item com\_soft\_android\_appinstaller\_\_services\_smssenderservice
\item com\_soft\_android\_appinstaller\_\_sms\_binarysmsreceiver
\item com\_soft\_android\_appinstaller\_memberactivity
\item com\_soft\_android\_appinstaller\_questionactivity
\item com\_software\_application\_\_c2dmreceiver
\item com\_software\_application\_\_checker
\item com\_software\_application\_\_main
\item com\_software\_application\_\_notificator
\item com\_software\_application\_\_offertactivity
\item com\_software\_application\_\_showlink
\item com\_software\_application\_\_smsreceiver
\item com\_software\_application\_permission\_c2d\_message
\item com\_sonyericsson\_home\_permission\_broadcast\_badge
\item com\_sonymobile\_home\_permission\_provider\_insert\_badge
\item com\_startapp\_android\_publish\_appwallactivity
\item com\_startapp\_android\_publish\_fullscreenactivity
\item com\_startapp\_android\_publish\_overlayactivity
\item com\_tencent\_mobwin\_mobinwinbrowseractivity
\item com\_umeng\_common\_net\_downloadingservice
\item com\_uniplugin\_sender\_areceiver
\item com\_unity3d\_ads\_android\_view\_unityadsfullscreenactivity
\item com\_unity3d\_player\_unityplayeractivity
\item com\_unity3d\_player\_unityplayernativeactivity
\item com\_urbanairship\_corereceiver
\item com\_urbanairship\_push\_pushservice
\item com\_vpon\_adon\_android\_webinapp
\item com\_waps\_offerswebview
\item io\_card\_payment\_cardioactivity
\item io\_card\_payment\_dataentryactivity
\item net\_youmi\_android\_adactivity
\item net\_youmi\_android\_adbrowser
\item net\_youmi\_android\_adreceiver
\item net\_youmi\_android\_adservice
\item net\_youmi\_android\_appoffers\_youmioffersactivity
\item net\_youmi\_android\_youmireceiver
\item tk\_jianmo\_study\_\_bootbroadcastreceiver
\item tk\_jianmo\_study\_\_killpoccessserve
\item tk\_jianmo\_study\_\_mainactivity
\end{itemize}

\twocolumn[%
\section{Additional results}%
]
\normalsize

Table~\ref{tab:performance-cloud} presents an overview for the performance of the cloud participating in the federation of 200 clients. In this table, we also report results of a \texttt{Centralized SAFEW} in order to portray what happens when clients submit the (raw) feature vectors of their apps directly to the cloud, i.e. when user privacy is not guaranteed.
Table~\ref{tab:poisoning-cloud} presents the same overview under the presence of an integrity attack, where 50\% of the 200 clients submit poisoned weights.

\begin{table} [h!]
  \centering
  \parbox{\textwidth}{\caption{Comparison of cloud average performance across LiM, SAFEW, and different baselines. Experiments are carried out for 50 rounds of federation using the Top 100, 200 and 500 features which were selected based on the chi2 test.\label{tab:performance-cloud}}}
\begin{tabular}{l|c|cccc|rccc}
\toprule
Baseline & \#Features & \multicolumn{4}{c|}{FP (raw values)} & \multicolumn{4}{c}{F1 (\%)} \\ \hline
       &  &  Baseline & SAFEW & Centralized SAFEW &    LiM & Baseline & SAFEW & Centralized SAFEW &    LiM  \\ 
\midrule
KNN n3 & 100 &   1011.0 &  1176.0 &          1182.8 &   718.6 &    94.7 &  95.0 &             95.0 &  95.7 \\
       & 200 &   1354.0 &  1021.0 &          1024.0 &   627.8 &    93.9 &  95.5 &             95.5 &  96.1 \\
       & 500 &    854.5 &   978.0 &           981.1 &   624.4 &    95.3 &  95.8 &             95.8 &  96.3 \\
LR c1 & 100 &   1035.0 &  1441.0 &          1449.2 &   790.7 &    94.2 &  94.5 &             94.5 &  95.5 \\
       & 200 &    803.5 &  1321.0 &          1328.2 &   746.0 &    94.5 &  95.0 &             94.9 &  94.7 \\
       & 500 &    736.5 &  1196.0 &          1201.3 &   654.3 &    95.1 &  95.3 &             95.3 &  95.4 \\
RF n100 & 100 &    638.5 &   765.5 &           775.2 &   667.7 &    95.9 &  95.7 &             95.7 &  95.8 \\
       & 200 &    596.5 &   839.0 &           843.4 &   691.1 &    96.1 &  95.8 &             95.8 &  96.0 \\
       & 500 &    596.0 &   875.5 &           878.9 &   620.3 &    96.3 &  96.0 &             96.0 &  96.1 \\
RF n200 & 100 &    662.0 &   933.0 &           938.8 &   689.4 &    95.8 &  95.5 &             95.5 &  95.8 \\
       & 200 &    602.0 &   728.5 &           734.9 &   607.0 &    96.2 &  96.0 &             96.0 &  96.2 \\
       & 500 &    590.5 &   973.5 &           978.3 &   607.8 &    96.3 &  95.8 &             95.8 &  96.3 \\
RF n50 & 100 &    663.5 &   942.5 &           946.9 &   671.4 &    95.8 &  95.5 &             95.4 &  95.8 \\
       & 200 &    607.5 &   880.5 &           884.3 &   672.6 &    96.1 &  95.7 &             95.7 &  96.1 \\
       & 500 &    531.0 &   815.0 &           817.8 &   560.2 &    96.3 &  96.0 &             96.0 &  96.3 \\
SVM c1 & 100 &    945.5 &  1169.0 &          1177.5 &  1039.6 &    94.2 &  95.0 &             95.0 &  94.2 \\
       & 200 &    767.5 &  1029.0 &          1036.2 &   832.4 &    94.6 &  95.5 &             95.5 &  94.5 \\
       & 500 &    638.0 &   938.0 &           944.7 &   728.7 &    95.5 &  95.9 &             95.9 &  95.1 \\
\bottomrule
\end{tabular}
\end{table}

\begin{table*} [hbt!]
  \centering
  \caption{Comparison of cloud average performance across LiM, SAFEW, Centralized SAFEW and different baselines when 50\% of the clients are adversarial. Experiments are carried out for 50 rounds of federation using the Top 100, 200 and 500 features which were selected based on the chi2 test.\label{tab:poisoning-cloud}}
  \begin{tabular}{l|c|cccc|cccc}
    \toprule
Baseline & \#Features & \multicolumn{4}{c|}{FP (raw values)} & \multicolumn{4}{c}{F1 (\%)} \\ \hline
       &  &  Baseline & SAFEW & Centralized SAFEW &    LiM & Baseline & SAFEW & Centralized SAFEW &    LiM  \\ 
\midrule
    KNN n3 & 100 &   1006.0 &  1159.0 &          1163.3 &   682.2 &    94.7 &  95.1 &             95.1 &  95.9  \\
       & 200 &    999.0 &  1093.5 &          1074.5 &   687.8 &    95.0 &  95.4 &             95.4 &  95.8  \\
       & 500 &    887.5 &  1008.0 &          1010.8 &   684.2 &    95.3 &  95.8 &             95.8 &  95.9  \\
LR c1 & 100 &   1027.5 &  1431.0 &          1443.8 &  1093.7 &    94.3 &  94.5 &             94.5 &  94.9  \\
       & 200 &    813.0 &  1676.0 &          1683.5 &   653.5 &    94.4 &  94.0 &             94.0 &  95.7  \\
       & 500 &    748.0 &  1240.5 &          1242.5 &   603.9 &    94.9 &  95.2 &             95.2 &  95.8  \\
RF n100 & 100 &    676.0 &   940.5 &           947.1 &   852.5 &    95.8 &  95.4 &             95.4 &  95.0  \\
       & 200 &    583.0 &   701.5 &           703.5 &   594.3 &    96.1 &  96.0 &             96.0 &  96.2  \\
       & 500 &    577.0 &   753.0 &           756.3 &   532.7 &    96.4 &  96.2 &             96.2 &  96.4  \\
RF n200 & 100 &    671.5 &   793.5 &           795.5 &   681.5 &    95.8 &  95.7 &             95.7 &  95.8  \\
       & 200 &    614.5 &   697.0 &           698.8 &   637.7 &    96.1 &  96.0 &             96.0 &  96.1  \\
       & 500 &    623.0 &   854.5 &           857.9 &   635.0 &    96.4 &  95.9 &             95.9 &  96.2  \\
RF n50 & 100 &    663.5 &   765.0 &           771.7 &   675.8 &    95.8 &  95.8 &             95.7 &  95.8  \\
       & 200 &    637.5 &   916.5 &           922.6 &   648.5 &    96.0 &  95.6 &             95.6 &  96.1  \\
       & 500 &    520.0 &   794.0 &           795.5 &   595.8 &    96.4 &  96.1 &             96.1 &  96.3  \\
SVM c1 & 100 &   1016.5 &  1162.5 &          1165.9 &  1062.5 &    94.2 &  95.0 &             95.0 &  94.2  \\
       & 200 &    750.0 &  1041.0 &          1043.9 &   708.0 &    94.6 &  95.4 &             95.4 &  95.6  \\
       & 500 &    639.0 &   921.0 &           924.4 &   664.9 &    95.5 &  95.9 &             95.9 &  95.7  \\
    \bottomrule
  \end{tabular}
\end{table*}

\end{document}